\documentstyle[12pt]{article}
\def\P{\Phi}
\def\p{\phi}

\def\m{\mu}
\def\r{\rho}

\def\n{\nu}

\def\a{\alpha}
\def\b{\beta}
\def\g{\gamma}

\def\pa{\partial}
\def\d{\delta}
\def\e{\epsilon}
\def\i{\bibitem}
\def\be{\begin{eqnarray}}
\def\ee{\end{eqnarray}}
\def\nn{\nonumber}
\def\f{\frac}

\begin{document}

\title
{\bf \LARGE Self-Duality of a Topologically Massive  Born-Infeld Theory}
\author{Prasanta K. Tripathy\thanks{email: prasanta@iopb.res.in}\\
and \\
Avinash Khare\thanks{email: khare@iopb.res.in} \\
\normalsize{\it 
Institute of Physics, Bhubaneswar 751005, India} }

\maketitle
\begin{abstract}
We consider self-duality in a $2+1$ dimensional gauge theory containing 
both the  Born-Infeld and the Chern-Simons terms. We introduce a Born-Infeld
inspired generalization of the Proca term and show that the corresponding
model is equivalent to the Born-Infeld-Chern-Simons model. 
\end{abstract}

\newpage
Many years ago Townsend et. al. studied self-duality in 
gauge theories in $4k - 1$ dimensions \cite{plich}. In particular,
in $2+1$ dimensions they considered 
the Proca equation for the massive gauge field:
\be
\pa^{\m}F_{\m\n} + m^2 A_{\n} = 0 ~,
\ee
where $F_{\m\n} = \pa_{\m}A_{\n} - \pa_{\n}A_{\m} $ .
As a consequence of the antisymmetry of the field strength, it follows 
from above that $\pa_{\m}A^{\m} = 0$, and hence there are two, 
independent, propagating modes of equal mass. They observed that any 
gauge field which is proportional to the dual of it's field strength
does satisfy the above equation.
In particular any gauge field which satisfy
\be
A_{\m} = \f{1}{2m} \e_{\m\n\r} F^{\n\r} ~,
\ee
is a solution of the second order Eq. (1). They called Eq. (2) as
the self-duality equation.  This equation propagates one massive 
mode instead of two and it  can be viewed as a square root of the 
second order Eq. (1). The self-dual Eq. (2) can be derived from 
the Lagrangian
\be
{\cal{L}}_{P} =   \f{1}{2} m^2 A_{\m} A^{\m} 
- \f{1}{4} m \e^{\m\n\r}A_{\m}F_{\n\r} ~.
\ee
It is straightforward to see that the above Lagrangian is not gauge 
invariant.  However, interestingly it was soon observed \cite{deser} 
that the above model is equivalent to gauge invariant, topologically 
massive, electrodynamics characterized by the Lagrangian \cite{schon,deser1}
\be
{\cal{L}}_M = - \f{1}{4}F_{\m\n} F^{\m\n} 
+ \f{1}{4} m \e^{\m\n\r}A_{\m}F_{\n\r} ~.
\ee
The corresponding field equation is 
\be
\pa_{\m}F^{\m\n} + \f{1}{2} m \e^{\n\a\b}F_{\a\b} = 0 ~,
\ee 
and following \cite{deser} it is easily shown that the field Eqs. (5) 
and (2) are equivalent. In fact, in  \cite{deser} the authors have
even shown the equivalence of the two Lagrangians ${\cal{L}}_P $
and ${\cal{L}}_M$ as given by Eqs. (3) and (4) respectively. 

Long back Born and Infeld proposed \cite{infeld} a nonlinear 
generalization to the Maxwell Lagrangian in order to cure the short 
distance divergence appearing in quantum electrodynamics. 
Recently it has attracted considerable attention both in field theory, 
because of it's remarkable form, as well as in string theory for it is the 
action which governs the gauge field dynamics of the D-branes \cite{born}. 
Because of its importance in the open string theory, Gibbons and Rasheed
studied various duality invariances of the Born-Infeld theory \cite{gibbons}.
In particluar, they have shown that the $SO(2)$ electric-magnetic duality 
rotation, that appears as a symmetry at the level of equations of motion 
in the Maxwell theory in four spacetime dimensions, also holds in the 
Born-Infeld theory. 
Because of the importance of duality in understanding various 
non-perturbative aspects of field theroy as well as string theory, 
the above results have been generalized to nonlinear theories with 
more then one Abelian gauge field, theories with interacting scalar 
fields as well as to the supersymmetric theories \cite{zumino}-\cite{aat} .
However most of the discussion about the duality invariance has been 
restricted to theories in four space time dimensions or more generally 
to the even dimensional theories.

On the other hand, several interesting generalizations of the self-dual 
Chern-Simons-Proca~model \cite{plich} and its equivalence \cite{deser}
with the three dimensional massive electrodynamics \cite{schon,deser1} 
has been studied in literature. Soon after the work of Deser and Jackiw, 
it has been realised that the self-duality can also occur in case both 
the Maxwell as well as the Proca term can simultaneously be incorporated 
in addition to the Chern-Simons term\cite{paul}. The above model has also
been used in the study of bosonization in higher dimensions 
\cite{fidel,robin}. Recently it has been shown that 
there exists a unified theory \cite{subir} from which the self-dual 
model\cite{plich}, the massive electrodynamics \cite{schon,deser1} 
as well as the Maxwell-Chern-Simons-Proca systems \cite{paul} can be 
recovered as special cases. However, to the best of our knowledge, the 
reuslts of Deser and Jackiw have not been generalized to the Born-Infeld 
theory.  The purpose of this note is to consider the generalization of 
this equivalence in case the Maxwell term is replaced by the celebrated 
Born-Infeld Lagrangian.  

Consider the Lagrangian 
\be
{\cal{L}}_{BI} = \b ^2\sqrt{1 - \f{1}{2\b ^2}F_{\m\n} F^{\m\n}} 
+ \f{1}{4} m \e^{\m\n\r}A_{\m}F_{\n\r} ~.
\ee
Here we have ignored an irrelevant constant factor proportional to square of
the Born-Infeld parameter $\b$ which does not contribute to the equation
of motion. This Lagrangian reduces to the topologically massive Lagrangian 
as given by Eq. (4) in the limit $\b \rightarrow \infty $ when the constant 
factor is taken into account. The corresponding field equation is 
\be
\pa_{\m}\left(\f{F^{\m\n}}{\sqrt{1 - \f{1}{2\b ^2}F_{\a\r} F^{\a\r}}}\right)
+\f{1}{2}m\e^{\n\a\r}F_{\a\r} = 0 ~.
\ee

The question one would like to ask is: 
what is the self-dual analogue of Eq. (2)? 
We  will now show that the corresponding ``generalized self-dual equation'' is 
\be
A_{\m} = \f{\e^{\m\n\r} 
F_{\n\r}}{ 2m \sqrt{1 - \f{1}{2\b ^2}F_{\m\n} F^{\m\n}}} ~.
\ee
Before we prove our assertion, let us note that Eq. (8) reduces 
to the self dual Eq.~(2) in the $\b \rightarrow \infty $ limit.
We call it the ``generalized self-dual equation'' because of it's 
similarity with the self-dual equation even though in the literature 
the term self-dual is usually used while dealing with  
the linearized equations. To show that Eq. (7) 
follows from Eq. (8), differentiate both sides of  
Eq. (8) and multiply by $\e^{\b\g\m}$. We get
\be
\e_{\b\g\m}\pa^{\g}A^{\m} = 
\f{1}{2m}\e_{\b\g\m}\e^{\m\n\r}\pa^{\g}
\left(\f{F_{\n\r}}{\sqrt{1 - \f{1}{2\b ^2}F_{\m\n} F^{\m\n}}}\right)  ~,
\ee
from which Eq. (7) follows.

Following  \cite{plich} it is worth enquiring if there is a 
corresponding Born-Infeld self-dual Lagrangian from which 
self-dual Eq. (8) will follow as a field equation. It is easily 
seen that such a Lagrangian is 
\be
{\cal{L}}_{BIP} =  \b ^2 \sqrt{1+\f{1}{\b^2}f_{\m}f^{\m}} 
- \f{1}{2m} \e^{\a\m\n} f_{\a} \pa_{\m}f_{\n} ~.
\ee 
Here we  have changed the notation from $A_{\m}$ to $f_{\m} = m A_{\m}$
in order to avoid confusion with the system described by Eq. (6).
The corresponding field equation is 
\be
\e^{\m\n\r} \pa_{\n}f_{\r} 
- \f{m f_{\m}}{\sqrt{1+\f{1}{\b^2}f_{\m}f^{\m}}} = 0 ~.
\ee
Let us now show that the above equation is equivalent to the 
generalized self-dual Eq. (8).
Taking the square of Eq. (11) we get
\be
\f{1}{2}f_{\m\n}f^{\m\n} = \f{m^2 f_{\m}f^{\m}}{1+\f{1}{\b^2}f_{\m}f^{\m}} ~,
\ee
which implies
\be
\sqrt{1 - \f{1}{2\b^2m^2}f_{\m\n}f^{\m\n}}
= \f{1}{\sqrt{1+\f{1}{\b^2}f_{\m}f^{\m}}} ~.
\ee
Substituting this again in Eq. (11) we get the generalized self-dual 
equation which can be identified with Eq. (8) after the field redefinition.
Thus both the theories described by Eqs (6) and (10) admit identical
solution for the self-dual gauge field. 

In fact the equivalence of the two theories is at a more basic level. 
In particular  we show that the corresponding Hamiltonians of the two 
theories are equivalent after the constraints are taken into account.
Let us start from the Born-Infeld Lagrangian (6). 
The corresponding canonical momentum is
\be
{\Pi ^i} &=& \f{\d{\cal{L}}_{BI}}{\d\dot{A}^i} \nn \\
&=& -\f{E^i}{\sqrt{1 - \f{1}{2\b^2}F_{\m\n}F^{\m\n}}} 
+ \f{1}{2}\e^{ij}A^j ~.
\ee
For convenience we define 
\be
R = \sqrt{1-\f{1}{2\b^2}F_{\m\n}F^{\m\n}} ~~, ~~~~
D^i = \f{E^i}{R}  ~. 
\ee
After some straightforward calculation we can write the expression for 
$R$ as 
\be
R = \sqrt{\f{1 - \f{1}{\b^2}B^2}{1 - \f{1}{\b^2}{\bf{D}}^2}} \, , 
\ee
where $B = F_{12}$ . \\
Now the Hamiltonian density is 
\be
{\cal{H}}_{BI} &=&  \Pi^i \dot{A}^i - {\cal{L}}_{BI} \nn \\
&=& -\f{E^{i}\dot{A}^i}{R} - \b^2 R \nn \\
&=& -\f{\b^2}{R}\left(1 - \f{1}{\b^2}B^2\right) \nn \\
&=& -\b^2 \sqrt{\left({1 - \f{1}{\b^2}{\bf{D}}^2}\right)
\left({1 - \f{1}{\b^2}B^2}\right)} ~.
\ee
The equal time commutation relation is  
\be
i \left[\Pi^{i}({\bf{r}}),A^{j}({\bf{r'}})\right] 
= \delta^{ij}\d{({\bf{r}}-{\bf{r'}})}
\ee
from which we can derive the following equal time commutators:
\be
i\left[D^{i}({\bf r}), D^{j}({\bf r'})\right] 
&=& m \e^{ij}\d({\bf r}-{\bf r'}) \\
i\left[D^{i}({\bf r}), B({\bf r'})\right] 
&=& -\e^{ij} \pa_{j}\d({\bf r - r'})  \\
i\left[B({\bf r}), B({\bf r'})\right] &=& 0 ~.
\ee 
These relations along with the Gauss law constraint
\be
\pa_{i}\Pi^{i} + \f{m}{2} B = 0 \, , 
\ee
can be solved in terms of a scalar field $\p$ as 
\be
D^{i} &=& - \e^{ij}\hat{\pa}_j\dot{\p} - m \hat{\pa}_i\p \\
B &=& \sqrt{-\nabla^2}\p ~,
\ee
where $\hat{\pa}_j = \pa_{j}/\sqrt{-\nabla^2}$ .

Now we consider the system described by Lagrangian (10). 
The conjugate momentum ${\Pi^i}_f $ is given by
\be
{\Pi^i}_{f} \equiv 
\f{\d{\cal{L}}_{BIP}}{\d\dot{f}^i} = - \f{1}{2m}\e^{ij}f^j \, ,
\ee
which implies the canonical commutation relation 
\be
i\left[f^i({\bf r}), f^j({\bf r'})\right]
= m\e^{ij}\d({\bf r - r'}) ~.
\ee
This along with the Gauss law constraint
\be
\f{f_0}{\sqrt{1+\f{1}{\b^2}f_{\m}f^{\m}}}
= -\f{1}{m}\e^{ij}\pa_{i}f_{j} \, ,
\ee
can also be solved in terms of the field $\p$ as 
\be
f^i = -\hat{\pa}_i\dot{\p} + m\e^{ij}\hat{\pa}_j\p \\
\f{f_0}{\sqrt{1+\f{1}{\b^2}f_{\m}f^{\m}}} = - \sqrt{-\nabla^2} \p  ~.
\ee 
Solving Eq. (29) for $f_0$ we get
\be
f_0 = - \sqrt{-\nabla^2} \p \left(\sqrt{\f{1 - \f{1}{\b^2}{\bf f}^2}
{1 - \f{1}{\b^2}(\sqrt{-\nabla^2} \p)^2}}\right) ~.
\ee
Now the Hamiltonian density is given by
\be
{\cal{H}}_f &=& {\Pi^i}_f\dot{f}^i - {\cal{L}}_{BIP} \nn \\
&=& - \f{1}{2m}\e^{ij}\left(f_0\pa_if_j + (\pa_if_j)f_o\right)
- \b^2\sqrt{1+\f{1}{\b^2}f_{\m}f^{\m}} \nn \\
&=& \f{f_0^2}{\sqrt{1+\f{1}{\b^2}f_{\m}f^{\m}}}
- \b^2 \sqrt{1+\f{1}{\b^2}f_{\m}f^{\m}} 
\ee 
where we have used the Gauss law to obtain the last step.
On using Eq. (30) the Hamiltonian density ${\cal{H}}_f$ takes the form
\be 
{\cal{H}}_f = - \b^2 \sqrt{\left(1 - \f{1}{\b^2}f_i^2\right)
\left(1 - \f{1}{\b^2}(\sqrt{-\nabla^2} \p)^2\right)} ~.
\ee
which is identical to the Hamiltonian density ${\cal{H}}_{BI}$ 
because of Eqs. (23), (24) and (28). 
This gives us the following identification of the fields $f_{\m}$ 
in terms of the field $A_{\m}$.
\be
f^{\m} = \f{F^{\m}}{R} ~,  
\ee
where 
\be
F^{\m} =  \e^{\m\a\b}\pa_{\a}A_{\b} \nn
\ee 
while $R$ is given by Eq. (15).

Before finishing this note it is 
worth enquiring if there is a  single ``generalized master Lagrangian''
from which the Lagrangians (6) and (10) follow. 
In this context it is worth pointing out that in  \cite{deser}, 
the authors have noted the common origin of 
the Lagrangians (3) and (4). In particular they have shown that these
Lagrangians follow from a single ``master Lagrangian''. 
Let us consider the Lagrangian 
\be
{\cal{L}}_{Mas} = \b^2\sqrt{1 + \f{1}{\b^2}f_{\m}f^{\m}}
- \e^{\m\a\b} f_{\m}\pa_{\a}A_{\b} 
+ \f{1}{2} m \e^{\m\a\b} A_{\m} \pa_{\a}A_{\b} ~.
\ee
Here we treat $f_{\m}$ and $A_{\m}$ as independent variables.
Varying this ``generalized master Lagrangian'' with respect 
to $f_{\m}$ and after a  little algebra we get Eq. (33). 
On eliminating $f_{\m}$ from ${\cal{L}}_{Mas}$ by 
using Eq. (33), we get the Lagrangian ${\cal{L}}_{BI}$ as given by 
Eq. (6). On the other hand, on varying Eq. (34) with respect to $A_{\m}$ we get 
\be
\e^{\m\a\b} \pa_{\a}f_{\b} = m F^{\m}  ~.
\ee
On using this to eliminate $A_{\m}$ from Lagrangian (34) gives the 
Lagrangian ${\cal{L}}_{BIP}$ as given by Eq. (10). Note that 
in the limit  $\b \rightarrow \infty $ the Lagrangian (34) reduces to 
the ``master Lagrangian'' of \cite{deser}.

To conclude, we have studied the generalized self-duality in
the topologically massive Born-Infeld theory and shown that the 
equivalence of Maxwell-Chern-Simons thoery \cite{schon,deser1} with 
the Chern-Simons-Proca theory \cite{plich} also holds in the nonlinear 
Born-Infeld theory. Here it is worth mentioning that the $2+1$ 
dimensional Born-Infeld action is the world volume action for 
D2-brane which can appear in the type $IIA$ superstring theory and 
can have Ramond-Ramond coupling  via the Chern-Simons term. More 
generally, the action for $n$ $Dp$-branes at small separation is 
descried by the Dirac-Born-Infeld action
\be
S_{dbi} = \int d^{p+1}\sigma 
Tr\left( e^{\P} \sqrt{-det(G + B + F)}\right)
\ee 
which can couple to the Ramond-Ramond background via the 
Chern-Simons term
\be
S_{cs} = \int_{p+1} Tr\left[e^{(B + F) \wedge C} \right]
\ee
(Here we have used the stringy notation, where $B$ is the Neveu-Schwarz
2-form pulled back on the world volume, $C$ is the Ramond-Ramond
field. $\P$ is the dilaton, $G$ the pull back of the metric on the world 
volume of the brane and $\sigma$ is the coordinate on it.)
It will be remarkable, if similar results, as given in our present work, 
can also hold for the above more genreal D-brane action for arbitrary 
$p$ in presence of the $B$-field.  
Clearly, further investigation is required to explore this point.

\end{document}